\documentclass[preprint, 5p, twocolumn, times]{elsarticle}

\usepackage{amsmath}
\usepackage{float}

\usepackage{lineno,hyperref}
\modulolinenumbers[5]

\journal{Ultramicroscopy}

\begin{document}
\begin{frontmatter}

\title{Event-based hyperspectral EELS: towards nanosecond temporal resolution}

\author[orsay]{Yves Auad\corref{mycorrespondingauthor}}
\cortext[mycorrespondingauthor]{Corresponding authors}
\ead{yves.maia-auad@universite-paris-saclay.fr}

\author[orsay]{Michael Walls}
\author[orsay]{Jean-Denis Blazit}
\author[orsay]{Odile Stéphan}
\author[orsay]{Luiz H. G. Tizei}
\author[orsay]{Mathieu Kociak}
\author[lille]{Francisco De la Peña}
\author[orsay]{Marcel Tencé\corref{mycorrespondingauthor}}
\ead{marcel.tence@universite-paris-saclay.fr}

\address[orsay]{Laboratoire des Physique des Solides, Université Paris Saclay, Orsay, France}
\address[lille]{Unité Matériaux et Transformations, Université de Lille, Lille, France}


\begin{abstract}
The acquisition of a hyperspectral image is nowadays a standard technique used in the scanning transmission electron microscope. It relates the spatial position of the electron probe to the spectral data associated with it. In the case of electron energy loss spectroscopy (EELS), frame-based hyperspectral acquisition is much slower than the achievable rastering time of the scan unit (SU), which sometimes leads to undesirable effects in the sample, such as electron irradiation damage, that goes unperceived during frame acquisition. In this work, we have developed an event-based hyperspectral EELS by using a Timepix3 application-specific integrated circuit detector with two supplementary time-to-digital (TDC) lines embedded. In such a system, electron events are characterized by their positional and temporal coordinates, but TDC events only by temporal ones. By sending reference signals from the SU to the TDC line, it is possible to reconstruct the entire spectral image with SU-limited scanning pixel dwell time and thus acquire, with no additional cost, a hyperspectral image at the same rate as that of a single channel detector, such as  annular dark-field. To exemplify the possibilities behind event-based hyperspectral EELS, we have studied the decomposition of calcite (CaCO$_3$) into calcium oxide (CaO) and carbon dioxide (CO$_2$) under the electron beam irradiation.
\end{abstract}

\begin{keyword}
\texttt{electron microscope; electron energy-loss spectroscopy; event-based; hybrid pixel direct detector; timepix3}
\end{keyword}
\end{frontmatter}


\section*{Introduction}

The scanning transmission electron microscope (STEM) works by rastering a focused electron beam on a sample. The image formation is usually performed by the single-channel annular dark field (ADF), bright field (BF), or annular bright-field (ABF) detectors. As the transmitted electrons carry spectral information from the sample, the focused electron probe makes STEM an interesting tool for performing electron energy loss spectroscopy (EELS) with high spatial resolution \cite{batson1993simultaneous, browning1993atomic, nelayah2007mapping, krivanek2010atom}. Data is usually acquired in the form of a hyperspectral image, a data cube indexed by one energy and two spatial coordinates.

One of the main concerns when performing EELS is that the energy-momentum transferred during the inelastic scattering of the electron may cause undesired effects in the sample, such as knock-on displacement, induced heating, and radiolysis \cite{egerton2011electron, pennycook2017impact}. Several approaches have been proposed to diminish them, such as custom scan paths and fast scans combined with data reconstruction algorithms \cite{zobelli2020spatial, stevens2018sub, trampert2018should, li2018compressed}. Although effective, these solutions are limited by the frame-based nature of the acquisition systems, which have a minimum acquisition time given by the readout time, typically of a few milliseconds for charge-coupled device (CCD) cameras, for example.

\begin{figure}[H]
    \centering
    \includegraphics[width=0.48\textwidth]{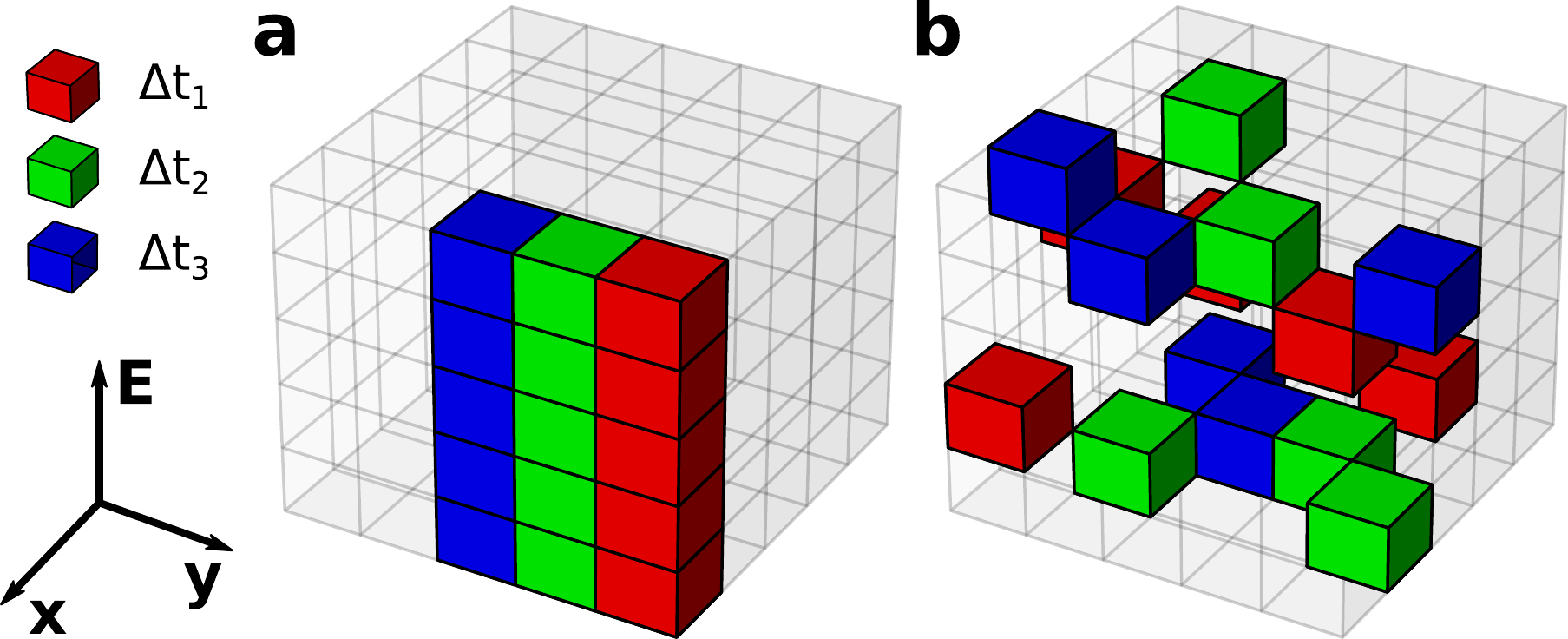}
    \caption{Comparison between frame-based and event-based hyperspectral acquisitions. (a) In the frame-based hyperspectral image reconstruction, the entire spectral dimension is acquired for each electron probe position. The minimum exposure time is given by the camera readout time, typically in the millisecond range for CCDs. (b) The event-based reconstruction places each electron in its corresponding data cube position when an electron hit is detected. Because of this, the electron beam can be rastered as fast as the time resolution of the event-based camera, typically in the nanosecond range. In both cases, the cube color code represents a typical acquisition time of a frame-based measurement ($\sim$ 1 ms). In such a time window, the scan unit can raster a great number of pixels.}
    \label{figEventBased}
\end{figure}

Up until now, frame-based detectors have been the usual solution for EELS acquisition. These count the number of electron hits in a given dwell time indiscriminately and thus the temporal information is limited by the spectrum acquisition time, as shown in Figure \ref{figEventBased}a. CCDs and complementary metal–oxide–semiconductor (CMOS) are the most widespread frame-based detectors for EELS \cite{strauss1987ccd, faruqi2005direct}. For both detectors, a scintillator and an array of optical fibers are typically used to convert the incident electrons into photons. These detectors have a variety of noise sources, such as dark and readout noises, and can dramatically degrade the spectral resolution due to the increased point-spread-function (PSF) imposed by the scintillator layer. A second kind of electron detection uses hybrid pixel detectors (HPDs), in which the sensor layer and the readout chip (also called application-specific integrated circuit or ASIC) are manufactured independently from each other. Multiple successive generations of ASICs led to the spread of HPDs in many different research subjects, such as space dosimetry \cite{stoffle2015timepix}, synchrotron source imaging \cite{ponchut2007photon, pennicard2014lambda}, X-Ray spectroscopies \cite{russo200818f, jakubek2009energy} and electron microscopy, including diffraction \cite{nederlof2013medipix, van2016ab}, imaging \cite{mcmullan2007electron, van2009medipix, krajnak2016pixelated, van2020sub} and EELS \cite{hart2017direct, goodge2020atomic}. One of the most successful ASICs, the Medipix3, introduces several improvements with respect to CCDs and CMOS for EELS acquisition. These include the practically zero readout noise, the improved PSF due to the direct electron detection and the readout time as low as $\sim$ 500 $\mu$s \cite{ballabriga2011medipix3}. Despite their improved acquisition speed, the problems related to frame-based acquisition persist because scanning pixel time in a STEM can go as low as tens of nanoseconds. This is much faster than the readout time of any commercially available frame-based detector.

A different concept of hyperspectral data acquisition for EELS can be defined when electrons are individually counted and can be unequivocally placed in the corresponding spectral and positional coordinates of the data cube. For example, one can consider a fast rastering electron beam with 0.5 $\mu$s pixel time. In such a time interval, for a probe with $\sim$ 50 pA only  $\sim$ 150 electrons would hit the sample, most of them falling in the zero-loss peak (ZLP). For a single-pixel acquisition, there would not be enough electrons to produce a usable EELS spectrum. However, continuously scanning and adding the electrons in an event-based fashion can lead to a meaningful reconstruction of the data cube. We show such a scheme in Figure \ref{figEventBased}b. The $\Delta t$ shown is a typical frame-based acquisition time ($\sim$ 1 ms). For the event-based acquisition, the SU rasters a great number of pixels within the time interval $\Delta t$ that would be needed to collect a spectrum in the frame-based approach. Of course, contrary to the frame-based approach, the arrival time of each of these hits is known with a precision much better than $\Delta t$. Also, during $\Delta t$, one acquires electron hits from different points of the data cube in space.  One must therefore relate a given electron hit with the corresponding probe position to construct a hyperspectral image. In this case, hyperspectral images can be acquired with very fast scanning pixel dwell time and thus synchronously with the normal ADF imaging without any performance penalty.


\begin{figure*}[t!]
    \centering
    \includegraphics[width=1.0\textwidth]{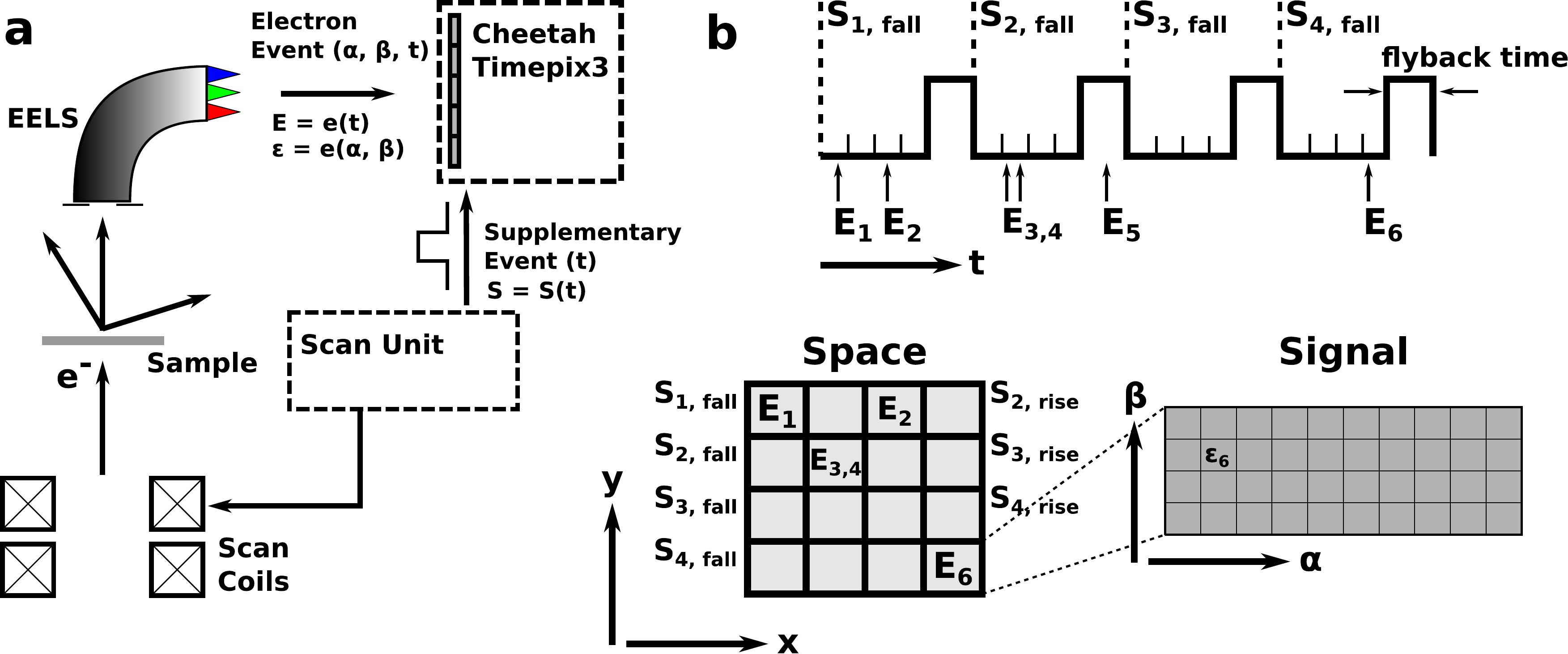}
    \caption{The hyperspectral data reconstruction process. (a) The scheme of the system used for data acquisition. The scan unit inputs temporal supplementary events, while individual electrons produce positional and temporal events. (b) We exemplify how the temporal information of both electrons and supplementary events can be used to arrange electrons in the reconstructed hyperspectral spatial data ($x$ and $y$). Detector-pixel address information ($\alpha$ and $\beta$) is used to determine the spectral information of each spatial pixel.}
    \label{fig1}
\end{figure*}

In this work, we demonstrate the implementation of this concept for EELS, similarly to what was recently demonstrated for event-driven 4D STEM acquisition \cite{jannis2022event}. Although there is a mention to a event-based hyperspectral image in the literature in the context of EELS-EDX coincidence experiments \cite{jannis2021coincidence}, the aforementioned study does not discuss the methodology nor the benefits of the time-resolved capability for time-dependent processes. We show that probe position and electron hit can be related to the temporal dimension by using an electron detector capable of outputting such information. We start by explaining details of the event-based hyperspectral EELS implementation, describing, in particular, the Timepix3, the direct electron detector used throughout this paper, and the related features of the used readout board, called SPIDR (Speedy Pixel Detector Readout), that allowed us to produce supplementary events from the Scan Unit (SU) superimposed on the data flow of the electron events. To illustrate the problem, we show an event-based hyperspectral acquisition using 120 ns pixel time sampled over 512 x 512 pixels. The last part of the paper is dedicated to the application of this system to follow the decomposition of calcite (CaCO$_3$) into calcium oxide (CaO) and gaseous carbon dioxide (CO$_2$) under electron beam irradiation.

\section*{Event-based hyperspectral EELS implementation}

To implement the event-based hyperspectral EELS, we have used a Timepix3 (TPX3) detector. In its first version, Timepix was a simple modification of Medipix2, allowing one to increment the pixel counter by clock ticks instead of the number of events since a reference clock was distributed on each one of its pixels. Timepix had thus the old functionality of counting hits but also the option of outputting either time of arrival (ToA) or time over threshold (ToT) values \cite{ballabriga2018asic}. The former measures the time elapsed until a hit is detected, while the latter measures the time the hit stays over the pixel signal threshold. 
Its successor, the ASIC TPX3, was the first real data-driven detector in the entire Medipix/Timepix family, as a pixel hit is responsible for triggering data output from the chip. A voltage-controlled oscillator running at 640 Mhz allows TPX3 to achieve a nominal temporal resolution of 1.5625 ns (called fine ToA) and, in contrast with the first Timepix generation, can simultaneously provide ToA and ToT \cite{llopart2007timepix, poikela2014timepix3}. When TPX3 is used in EELS, therefore, we have access to each electron's positional coordinates (the dispersive and non-dispersive directions) and the temporal coordinates, represented here by both ToA and ToT. To reconstruct the hyperspectral image, one must find a way to correlate the temporal information of the electron events with the electron probe position. One approach is to feed the SU reference clock signal into TPX3, which would require flexible and programmable SUs and TPX3 control boards. Our solution is to create supplementary events in the TPX3 data flows, effectively having two distinct kinds of events: one linked to individual electrons and another to reference timestamps of the microscope probe position. 

For the development of our application, we have used the TXP3 solution by Amsterdam Scientific Instruments (ASI), called Cheetah, which includes the SPIDR board \cite{van2017spidr} and the control software. Our detector consists of four 256x256 chips mounted linearly adjacent to each other to form a 256x1024-pixel array. In the following, the dispersive direction of the detector is denoted as $\alpha$ and the non-dispersive $\beta$. Also, the Cheetah provides us with two supplementary input time-to-digital converter (TDC) lines that run the same clock as the 40 Mhz reference clock and can reach a nominal temporal resolution of $\sim$ 260 ps. The SPIDR can detect TTL-based rising and falling edges in the TDC and includes them in the data flow in the same way it is done for electron events. We have used a custom-made SU solution that is based on a 25 Mhz clock and can scan as fast as 40 ns per pixel \cite{zobelli2020spatial}. To synchronize the SU and the SPIDR clocks, the SU sends reference signals (what we call supplementary events) to the Cheetah, as demonstrated in Figure \ref{fig1}a. They contain only timestamps and can be represented by any input signal that can be used to unequivocally determine the electron probe position $(x,~y)$. Although theoretically one could use a single signal indicating the start of the rastering, sending periodic reference signals allows to correct clock drift, which is especially important for long ($>$10 s) acquisition times. In our case, we have used the beginning of a new scan row ($y$ direction) as a trigger falling edge, while the end of a line is represented by a rising edge. The difference between a falling and a rising edge is the scanning flyback time setting.

The complete hyperspectral reconstruction principle is shown in Figure \ref{fig1}b, which depicts the timeline of the occurring events. For clarity, electron events $e_{n}$ are further subdivided into $E_n = e_{n}(t)$ and $\varepsilon_n = e_{n}(\alpha, \beta)$ to explicitly indicate what information we have used in each step of the reconstruction. As the received supplementary event S(t) relates to the beginning of a new scan row, the number of columns ($x$ direction) must be known by the software. This value is used as the number of time bins between a supplementary falling event and a successive rising event, as shown at the top in Figure \ref{fig1}b. As an example, we can see that electrons $E_1$ and $E_2$ are in the same row because they are both after $S_{1, fall}$ but are in different columns because they are in different time bins within the scan row. It is important to note that electron placement in the hyperspectral spatial pixel ($x$ and $y$) is only dependent on time. The pixel address of the electron event, $\varepsilon_n$, is only used to form the hyperspectral signal ($\alpha$ and $\beta$), as shown in Figure \ref{fig1}b at the bottom left by $\varepsilon_6$. Additionally, the rising edge trigger input by the SU indicates the end of a scan row, meaning the high digital signal corresponds to the flyback of the electron probe; any electron event that arrives during this time interval is rejected, as illustrated by $E_5$. As a final remark, it is important to clarify that the time $t$ in $E_n = e_{n}(t)$ is simply the electron ToA corrected by the fine ToA, having a nominal temporal resolution of 1.5625 ns. Note also that the multiple electron-hole pairs created by a single impinging electron create multiple detector hits, called clusters. To circumvent the problem of multiple event counting due to clusters, a cluster-correction algorithm was implemented in our application. It must use both the temporal and the spatial information of adjacent electron hits to be effective and is explained in detail later in this work.

\begin{figure*}[t]
    \centering
    \includegraphics[width=1.0\textwidth]{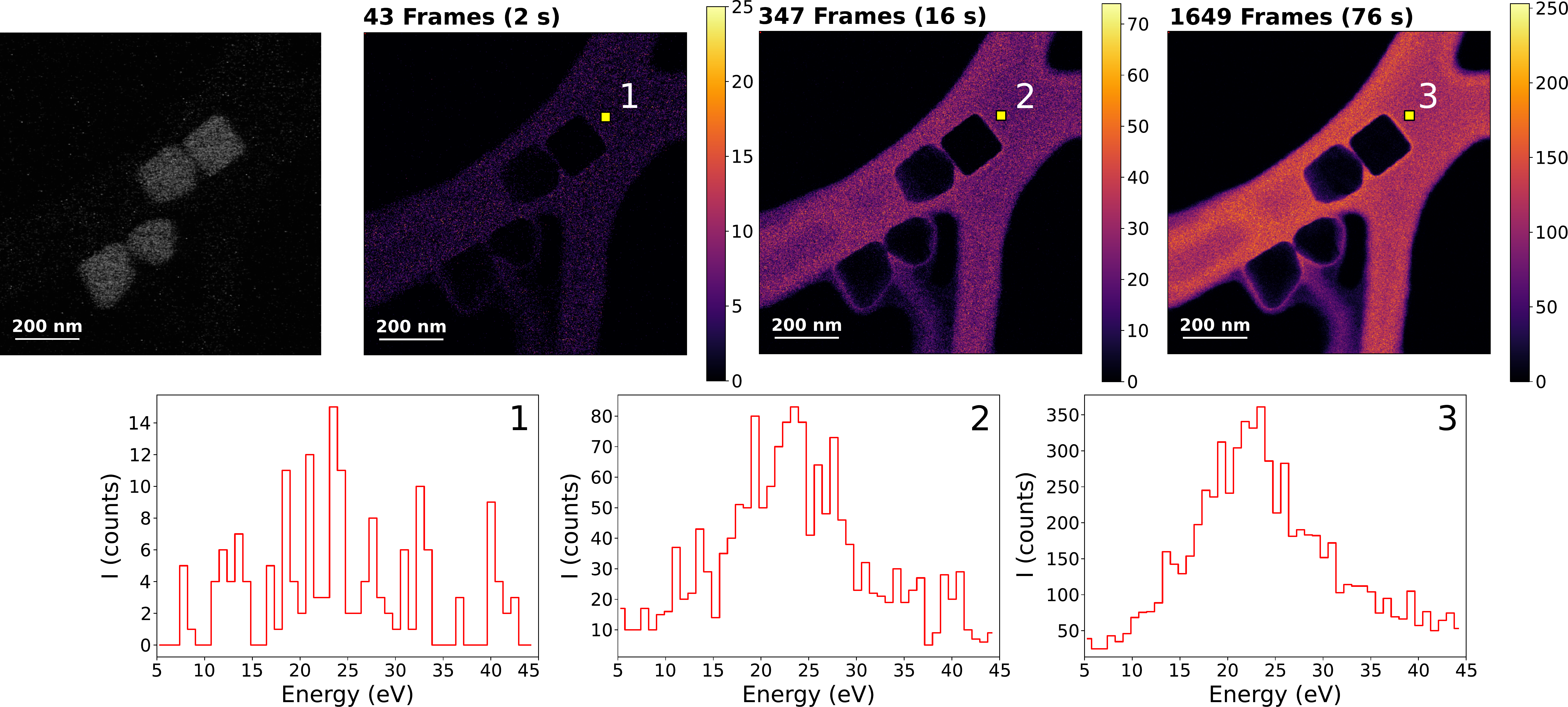}
    \caption{Energy-filtered hyperspectral image containing 512 x 512 pixels and using 120 ns pixel time between 5 eV and 45 eV for a sample of silver nano-cubes drop-casted over a thin amorphous carbon film. The ADF (top left) and three images (1, 2, and 3, at the top) and the corresponding spectra acquired inside the highlighted yellow square (8 x 8 pixels) taken after 64, 508, and 2415 complete ADF frames show the time evolution and the event-based nature of hyperspectral data formation.}
    \label{figCuveOverCarbon}
\end{figure*}

We have also developed a live acquisition program coded in Rust \cite{auad2022tp3tools} capable of translating the received events by the TPX3 to a variety of outputs, including the hyperspectral image illustrated in Figure \ref{fig1}. The software can be controlled in a user interface plugin \cite{auad2022nslumiere} developed for the Nionswift software \cite{meyer2019nion}, which is also used for data acquisition. Other software features include the acquisition of single spectra, which uses the period of a TTL signal in the TDC line to determine the spectrum dwell time. Both the live-processing and the plugin are open-source and are available to the community under MIT licensing. The processed data is transferred by transmission control protocol (TCP) using a 10-Gbit optical fiber from the dedicated processing computer to the client computer. For a single spectrum, data is transferred in its entirety (1024-sized array for fully-binned measurements and 1024x256-sized for image measurements) with configurable bit depth to accommodate a high range of acquisition times. For the hyperspectral image, one must note that for a 512 x 512 image with 120 ns pixel time, an entire 512 x 512 x 1024 hyperspectrum is simultaneously reconstructed with the ADF, although very sparse. The transfer rate would need to be $\sim$ 140 Gbit/s (using a 16-bit integer) which is much higher than the transfer limit of the 10-Gbit Ethernet. In such cases, data can be transferred more compactly by sending a list of indices to be incremented in the datacube. As an example, for a hyperspectrum containing 64 x 64 spatial pixels, and considering the 1024 pixels in the detector row, these indices must be between 0 and 4194304.


Figure \ref{figCuveOverCarbon} shows an initial example of a live hyperspectral reconstruction. Pixel dwell time was kept at 120 ns in a 512 x 512 spatial sampling with a current at the sample of approximately 8 pA from a region of approximately 1.0 $\mu$m$^2$. The flyback time is measured as $\sim$ 28.5 $\mu$s by the TDC and thus a single frame takes approximately 46.1 ms to acquire. The sample contains some silver nano-cubes drop-cast onto a thin film of amorphous carbon. At the top, the ADF image obtained during data acquisition and three snapshots for different accumulation times of the energy-filtered hyperspectrum between 5 eV and 45 eV, comprising the strong carbon plasmon resonance peaked at approximately 22 eV. At the bottom, we display the spectrum for the 8 x 8 pixel cell highlighted by the yellow square. In the first 2 s of acquisition, 43 complete ADF frames are accumulated and a minimal contrast shows up in the energy-filtered image. After 16 s and 76 s of acquisition (corresponding to, respectively, 347 and 1649 frames), the contrast is greater and the plasmon resonance is much more distinguishable.






\section*{Study of calcite decomposition}

In order to demonstrate our event-based hyperspectral image, we have used a calcite (CaCO$_3$) sample and explored its well-known transformation to calcium oxide (CaO) and carbon dioxide (CO$_2$) under the electron beam irradiation (CaCO$_3$ $\xrightarrow{e^-}$ CaO + CO$_2$) \cite{Tence1989, Walls1989, golla2014situ}. The experiment was performed in a Vacuum Generators HB501 at 100 kV equipped with an LN$_2$ cold stage that stays at approximately 150 K. The acquired data had 4 $\mu$s pixel time with the 32 nm x 32 nm region sampled by 32 x 32 pixels. The convergence angle was 15 mrad and a collection aperture of $\sim$ 2 mrad was used to have both an improved spectral resolution and to produce a non-saturated EELS dataset. The electron spectrometer was set to a low dispersion of $\sim$ 0.445 eV/pixel to monitor simultaneously the low loss region, the carbon K edge, and the calcium L$_{2, 3}$ edges. In these conditions, one ADF image, and therefore one hyperspectral image, is completed every $\sim$ 5 ms. Such a rate is comparable with that of a single energy-filtered transmission electron microscope (EFTEM) image, although in the present case the whole spectral range is gathered. The collected signal is extremely low at these rates, as the dwell time for the acquisition of each pixel's spectrum is $\sim 4~\mu s$, and some time-binning is needed for interpreting the data. Therefore, the total of 93 s of the acquisition was sliced into 232 hyperspectral images with intervals of 400 ms, which corresponds to roughly 80 complete ADF frames and an exposure time per pixel of 320 $\mu$s. As we shall see, this temporal sampling is enough to unveil the calcite decomposition dynamics in the low-loss energy range. Data analysis in this work was done using the Hyperspy package \cite{Hyperspy}. 

\begin{figure}[t]
    \centering
    \includegraphics[width=0.42\textwidth]{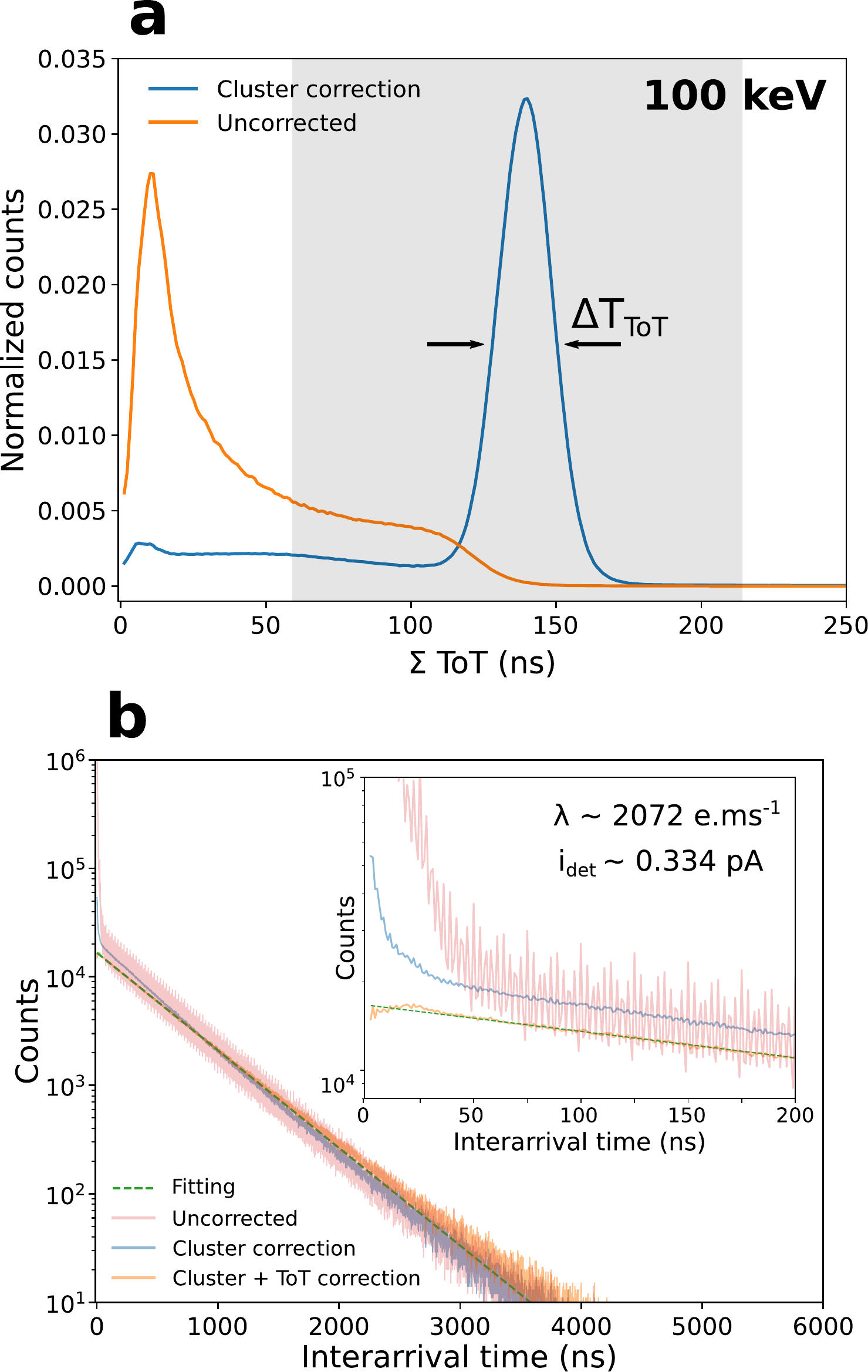}
    \caption{Impact of the cluster-correction algorithm in the EELS data. (a) The normalized frequency of the summed ToT before and after cluster correction. (b) The electrons inter-arrival times for the uncorrected, cluster-corrected and cluster+ToT-corrected data. Fitting was performed for the latter and provided an expected number of occurrences of $\lambda$ $\sim$ 2072 electrons.ms$^{-1}$, corresponding to a current in the detector of $i_{det}$ $\sim$ 0.334 pA.}
    \label{fig2}
\end{figure}

Before examining the data set, we used a custom-developed algorithm to identify and treat clusters from our hyperspectral time slices. To do so, both ToA and the pixel hit position are used: the set of pixels within a single cluster is counted as a single event carrying the average ToA and pixel impact position. A new cluster is created if the next electron event has a ToA superior to the previous one by $>$ 200 ns or if the pixel distance is $>$ 2 pixels in any of the $\alpha$ or $\beta$ directions independently (see Supplementary Material (SM) for further details on different parameters). Figure \ref{fig2} shows the impact of cluster treatment on our EELS hyperspectral data. In Figure \ref{fig2}a, we have plotted the histogram of the ToT for all pixel hits before the cluster correction (orange curve) and the histogram of the sum of the ToT of all the pixel hits
that belong to a single cluster (blue curve). A Gaussian fit to the distinct peak shown in the cluster-corrected data gives us an average value of $\sim$ 139.13 ns and an equivalent full-width-half-maximum (fwhm) $\Delta$T$_{ToT}$ = 22.65 ns, which is under the clock tick of 25 ns. In such a case, ToT-based spectroscopy has a resolution of approximately 16.18 keV and hence is difficult in the typical EELS range ($<$ 1 keV). In Figure \ref{fig2}b, we have plotted the time difference between consecutive events, called inter-arrival times (ITs), for the same electrons as in Figure \ref{fig2}a. A consequence of the independence of the events in a Poisson process is that the number of events as a function of the observed IT follows an exponential decay $e^{-\lambda t}$, where $\lambda$ is the expected rate of occurrences in the Poisson process. The uncorrected curve (light red) seems to be properly following an exponential decay for ITs longer than 100 ns but has a steep increase of approximately two orders of magnitude for ITs shorter than 50 ns. Additionally, the uncorrected curve presents oscillations in the observed ITs, which is also a consequence of the multiple detected hits per electron and the inability to determine the proper effective electron hit time.  After cluster-correction (light blue curve), the curve approximates to an exponential behavior for shorter times, despite a still visible deviation for ITs $<$ 25 ns. Additionally, we also show (light orange curve) the IT for the electrons in which their cluster total ToT is between 60 and 220 (gray rectangle in Figure \ref{fig2}a), which follows a much-closer exponential behavior for short ITs. As it is discussed in the SM, identified clusters with small total ToT are primarily formed in between Timepix3 chips and thus might be subjected to a different cluster formation dynamics. Finally, the current in the detector estimated by the number of hits after cluster + ToT correction is 0.322 pA. The fitting result (dashed line) gives $\lambda$ $\sim$ 2072 electrons.ms$^{-1}$, which corresponds to a current of $\sim$ 0.334 pA and agrees within 96\% with the electron hit estimate. Note that the Poisson statistics of the electrons are indicative of the non-saturated regime of electron detection.




\begin{figure}[H]
    \centering
    \includegraphics[width=0.42\textwidth]{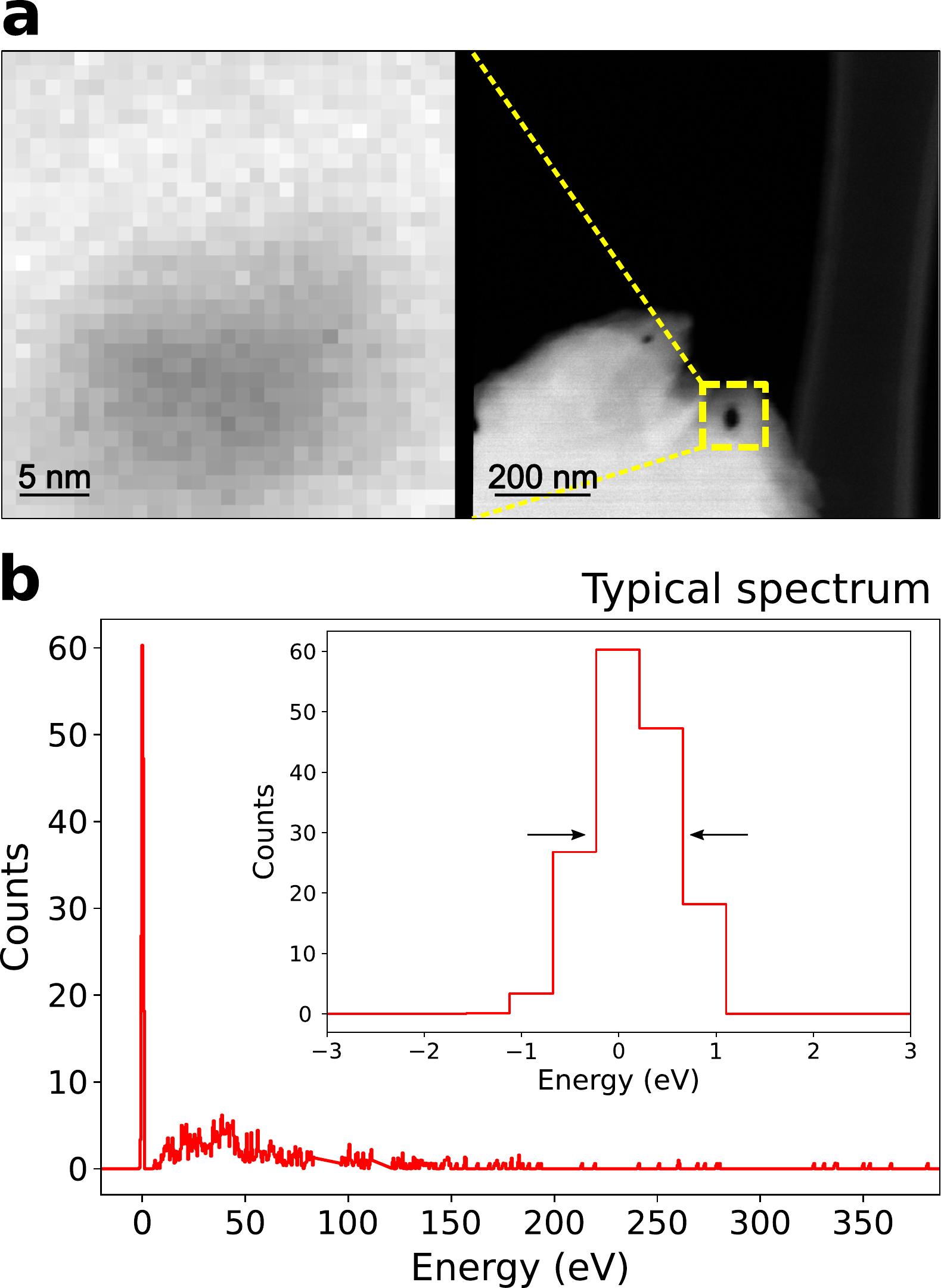}
    \caption{Typical acquisition conditions. (a) ADF images at approximately 25 s of acquisition time (left) and at the end of the acquisition after 93 s (right). (b) Typical EELS spectrum for a single pixel in a single time slice, showing a fwhm of 2 pixels.}
    \label{fig3}
\end{figure}

\begin{figure*}[ht]
    \centering
    \includegraphics[width=1.0\textwidth]{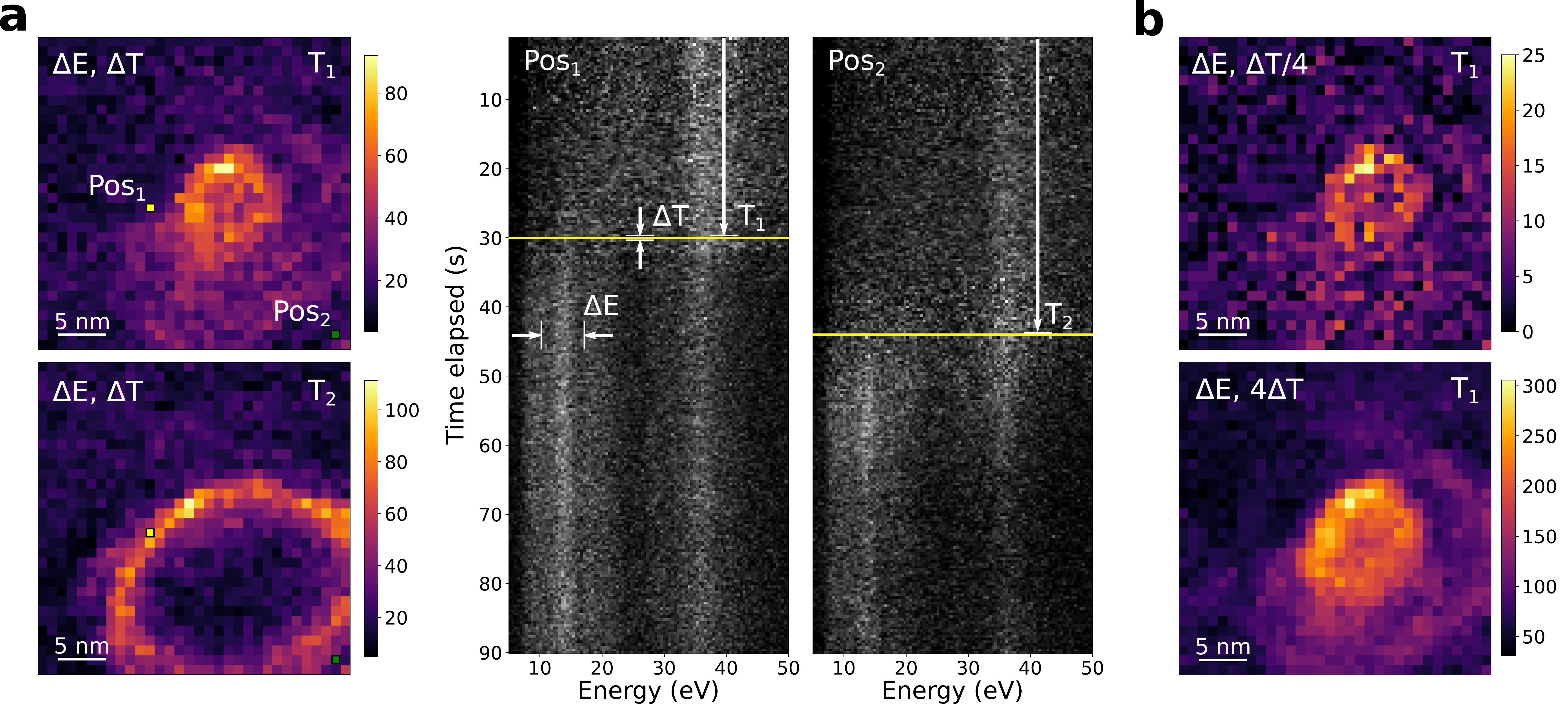}
    \caption{Hyperspectral EELS results for the calcite decomposition in the low-loss energy range. (a) Two energy-filtered snapshots centered at $E$ = 13 eV accumulated in the energy internal $\Delta E$ for the time slices at $T_1$ = 30 s and $T_2$ = 45 s summed over the time interval $\Delta T$ = 400 ms. The time evolution for two pixels (Pos$_1$ and Pos$_2$) is also shown. (b) Similar to the snapshots in (a), but for a time interval of 100 ms (top) and 1600 ms (bottom).}
    \label{fig4}
\end{figure*}

\begin{figure}[b!]
    \centering
    \includegraphics[width=0.48\textwidth]{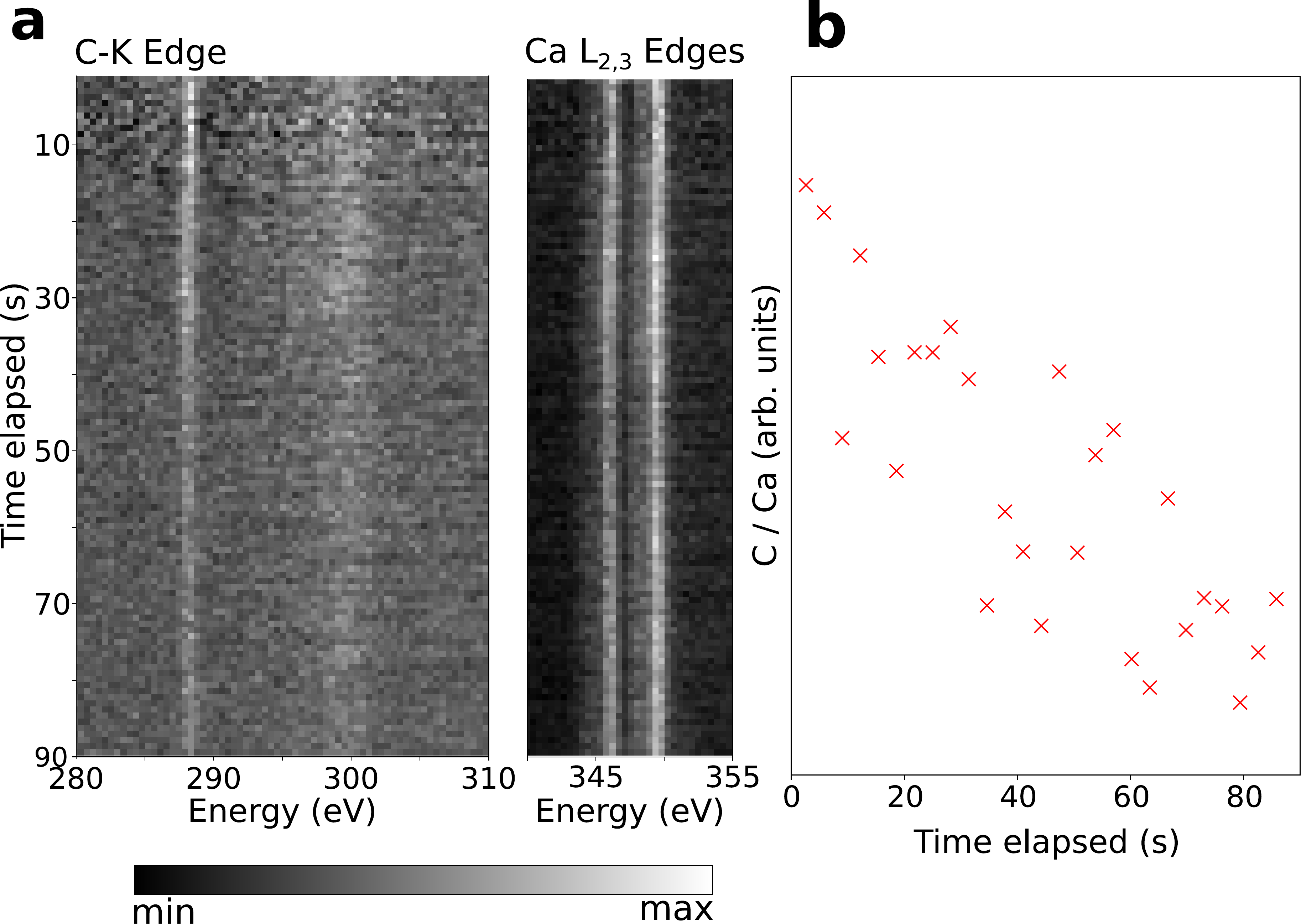}
    \caption{Hyperspectral EELS results for the calcite decomposition in the core-loss energy range for the entire sample region. (a) Time evolution around the C-K edge at the left while, at the right, we display around the calcium L$_{2,3}$ edges. (b) The ratio between C-K edge (286 eV - 305 eV) and the calcium-L$_{2,3}$ (343.5 eV - 351.0 eV). The diminishing proportion indicates that despite sample mass-loss, carbon content is reducing with respect to calcium.}
    \label{fig5}
\end{figure}

Figure \ref{fig3}a displays one snapshot of the ADF at $\sim$ 25 s of acquisition time (left), which already shows a contrast due to the accumulated sample damage. The ADF at the right shows a higher field-of-view image after the entire 93 s of acquisition. Figure \ref{fig3}b shows a typical single-pixel spectrum in one time slice (320 $\mu$s pixel exposure time), displaying a ZLP with a maximum number of $\sim$ 60 electron hits and a fwhm of 2 pixels, a consequence of the improved point-spread-function of direct electron detectors \cite{hart2017direct}.

Figure \ref{fig4}a shows a few results from the time-resolved hyperspectrum after running the cluster-correction algorithm in the data set. Two energy-filtered images centered at the plasmon resonance feature at $\sim$ 13 eV, indicated as $\Delta E$ and associated with the CaO formation \cite{Walls1989, golla2014situ}, are shown at the left for two distinct times ($T_1$ and $T_2$) within the same time interval $\Delta T$ = 400 ms and thus depicts the CaO formation dynamics within $\pm$ 200 ms time resolution. The EELS spectra as a function of the total elapsed time for the pixel Pos$_1$ (yellow square) and Pos$_2$ (green square) are shown at the right. Note how Pos$_2$ is farther from where CaO formation starts and hence the transformation is triggered at a later time than at Pos$_1$. There is a clear transformation in the low-loss spectra, most notably around the aforementioned resonance at $\Delta E$, successfully captured by the time-binning chosen. In Figure \ref{fig4}b, we show a similar energy-filtered snapshot, but with time intervals of 100 ms and 1600 ms, demonstrating that the time-binning value can be arbitrarily picked as long as it is a multiple of a single scan image. 

In Figure \ref{fig5}a, we show similar spectra for the core-loss energy range around the Carbon-K edge and the Calcium L$_{2,3}$ edges for the entire sample region.
In Figure \ref{fig5}b, we display the sum of the normalized signal between 286 eV and 305 eV (comprising thus the C-K edge) divided by the signal between 343.5 eV and 351.0 eV (Ca L$_{2,3}$). To have a better signal-to-noise ratio (SNR), time slices were binned by a factor of 8 (and thus have a time resolution of $\sim$ 3.2 s). The smaller proportion of carbon with respect to calcium over time suggests the calcite decomposition is happening and, consequently, carbon content is reducing in the system. 

\begin{figure}[H]
    \centering
    \includegraphics[width=0.4\textwidth]{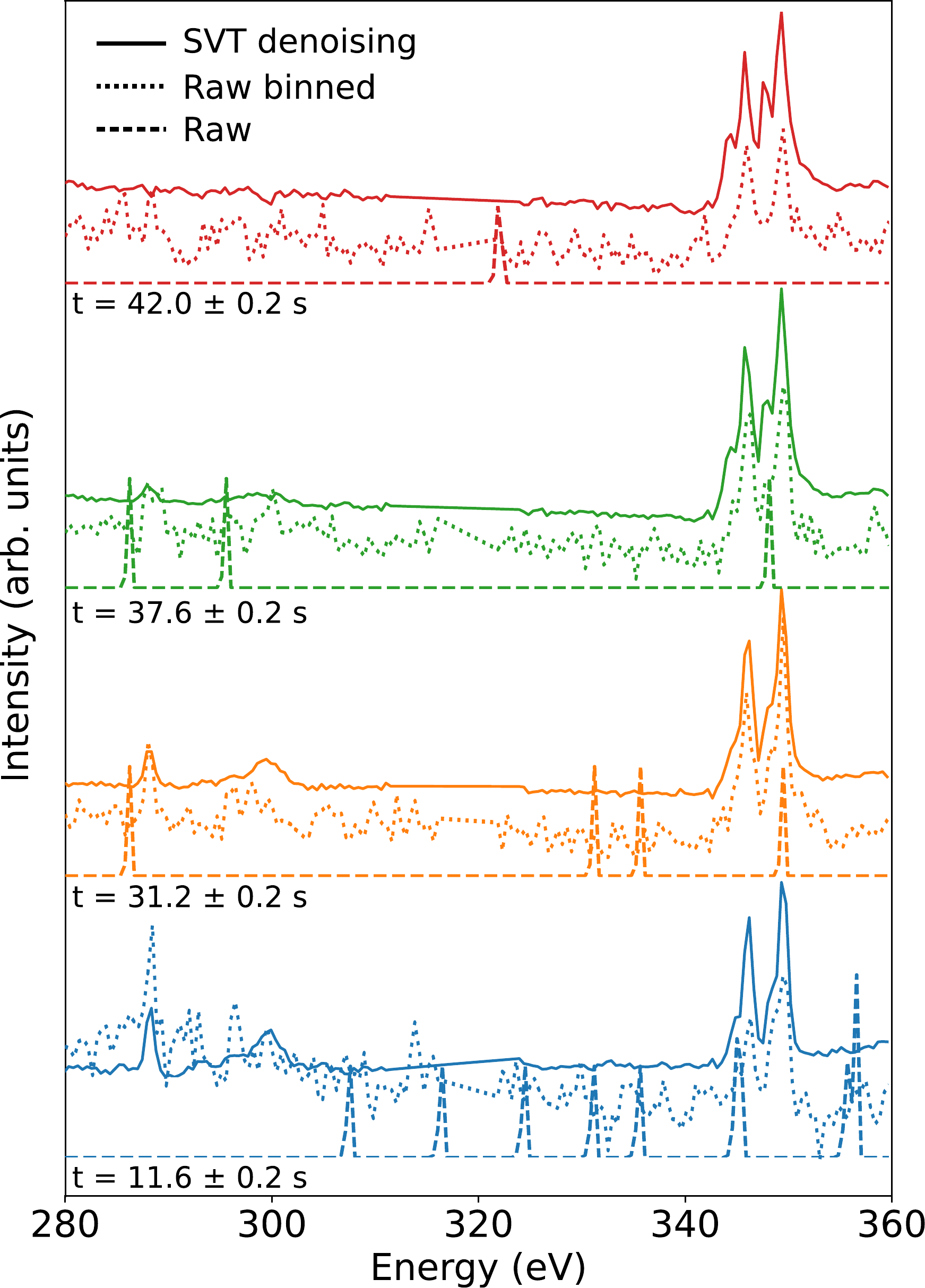}
    \caption{Set of spectra for a single pixel close to Pos$_1$ for four different times after SVT denoising (solid curve), for the pixel raw data associated (dashed curve), and for the 32x32 binned raw data (dotted curve). Carbon content is progressively reduced, while the crystal field splitting, associated with the Ca-O bonds, increases. Time slices are within the unbinned time interval of 400 ms and exposure time, per pixel, is 320 $\mu$s.}
    \label{fig6}
\end{figure}

To extract more spectral information from the calcite decomposition dynamics, one could further increase the time interval of the hyperspectral slices, sacrificing time resolution for more signal per spectrum. More interesting, however, is to perform a low-rank approximation, such as singular value thresholding (SVT), a.k.a. PCA (principal component analysis), which can increase the signal-to-noise ratio \cite{arenal2008extending, de2011mapping} without sacrificing time and spatial resolution. Figure \ref{fig6} shows the result of SVT of the dataset with 3 components for a single spatial position close to the yellow square (Pos$_1$) highlighted in Figure \ref{fig4}. The time evolution shows the progressive reduction of the carbon content, followed by a more and more pronounced crystal field splitting of the $t_{2g}$ and $e_g$ peaks in the Ca L$_{2, 3}$ edge due to the undistorted octahedral symmetry and change in length of the Ca-O bonds in CaO compared to CaCO$_3$ \cite{golla2014situ}. The SVT was performed with the ZLP and the pixels close to the chip edges masked. The raw spectra associated are also shown in Figure \ref{fig6} by the dashed superimposed curve for the same pixel as the SVT data, and by the dotted curve for the spatially binned 32x32 spectrum, which highlights the impressive potential of SVT denoising for such low-signal time-resolved datasets. Finally, note that although the time slice interval has the unbinned value of 400 ms, single-pixel exposure time is $\sim$ 320 $\mu$s. 

\section*{Conclusions and perspectives}
In conclusion, we have presented the acquisition of a hyperspectrum with scanning speed limited by the SU rastering time instead of the detection system. We have used a commercial event-based TPX3 solution along with an also commercially available custom-made scan engine \cite{Liskamm} in which external events from the SU add timestamps in the electron data flow that can be later used to retrieve the electron probe position. For this reason, we refer to our approach as event-based hyperspectral EELS. All the developed software is available to the community, including the live data processing \cite{auad2022tp3tools} and the interface plugin \cite{auad2022nslumiere}, and thus any SU capable of outputting the scan clock signal along with the Cheetah solution could be used to reproduce this work. To demonstrate our system capabilities, we have given as an example the decomposition of calcite into CaO and CO$_2$ under electron beam irradiation. After cluster correction and ToT correction, electron arrival times follow a Poisson distribution, which shows both the well-known statistics of the electron emission in a cold field emission gun (cFEG) and the non-saturated regime of the data acquisition. In principle, hyperspectral images can be acquired with pixel times as low as 1.5625 ns (nominal temporal resolution of TPX3), although reaching this scan rate would need further TPX3 calibrations \cite{bergmann20173d, pitters2019time} that are irrelevant for our minimum rastering time of 40 ns. In TPX3, data can be saturated by the pixel dead time, by the column readout scheme, and by the detector maximum throughput. As the ZLP is focused in a single detector column, applications in which a meaningful ZLP intensity is required might be restricted to detector currents up to 1-2 pA, although tilting the detector/electron beam or custom pixel-line masking might alleviate this problem. The maximum detector throughput corresponds to currents up to 10-15 pA, which can also limit the detector applications. Improvements in the near future for all these aspects are expected with the new Timepix4 detector \cite{campbell2016towards}. Time-resolved data is shown for the CaO formation in the low-loss and the core-loss energy range. For the latter, we have achieved a single-pixel spectrum after performing signal decomposition in the hyperspectral slices. We believe that event-based EELS will become increasingly available in the microscopy community. They will effectively tackle several important problems that require both nanometric spectral resolution and nanosecond time resolution. These include optical microresonators \cite{auad2022unveiling, henke2021integrated, muller2021broadband} thanks to their long-lived excitations, and accessing the chemistry of electron-irradiation sensitive materials like graphene oxide \cite{tararan2016revisiting}. Additionally, the setup described in this work can be used to easily reconstruct photon-electron coincidence hyperspectral images, recently demonstrated \cite{varkentina2022cathodoluminescence}.


\section*{Data availability}
The raw data used in this paper was made available in Zenodo \cite{auad_yves_2021_5552559}. The live processing software developed and the Nionswift plugin interface in this work is also available under MIT license \cite{auad2022tp3tools, auad2022nslumiere}.




\section*{Acknowledgements}
The present project has received funding from the  European  Union’s  Horizon  2020  research and nnovation programme  undergrant  agreement  No  823717  (ESTEEM3)  and  101017720  (EBEAM). We thank Marta de Frutos for discussions on EELS data analysis. Amsterdam Scientific Instruments (ASI) is also acknowledged for many fruitful technical discussions.

\bibliography{newbiblio.bib}

\setcounter{figure}{0}
\makeatletter 
\renewcommand{\thefigure}{S\@arabic\c@figure}
\makeatother
\onecolumn

\clearpage
\begin{large}
\textbf{Supplementary Material for Event-based hyperspectral EELS: towards nanosecond temporal resolution}
\end{large}

\section*{Clock drift between the SU and the TPX3 readout board}

To understand how often the TPX3 and the SU clocks must be corrected, we have analyzed how many clock ticks, in units of the fine TDC bin of $\sim$ 260 ps, have been counted between successive scan rows concerning the main hyperspectral data of the manuscript text (Figure \ref{fig4}). For the more than 590000 scan rows analyzed, 76.7\% of them have the same number of clock ticks between successive lines, which is also the reference value for placing the electrons, as discussed in Figure \ref{fig1}. 18.2\% of the scan rows are shifted by one TDC clock tick. The other 5.1\% has a maximum shift of 2 clock ticks. The average clock drift is $\sim$ 0.23 ticks per every 601500 ticks of the whole scan line, or 1 cycle every 2615217 ticks. In the present case, this corresponds to $\sim$ 60 ps per line and hence an average drift, per frame, of $\sim$ 1.9 ns, much smaller than the pixel dwell time of 4 $\mu$s.

For the carbon membrane, considering the same logic applies, the drift per line would be $\sim$ 34 ps, or $\sim$ 17.4 ns per frame. This value is almost 8 times smaller than the pixel dwell time of 120 ns, meaning that, in the present experimental conditions, clock drift correction can be done either by each new scan row or each new can frame.

\begin{figure}[H]
    \centering
    \includegraphics[width=0.55\textwidth]{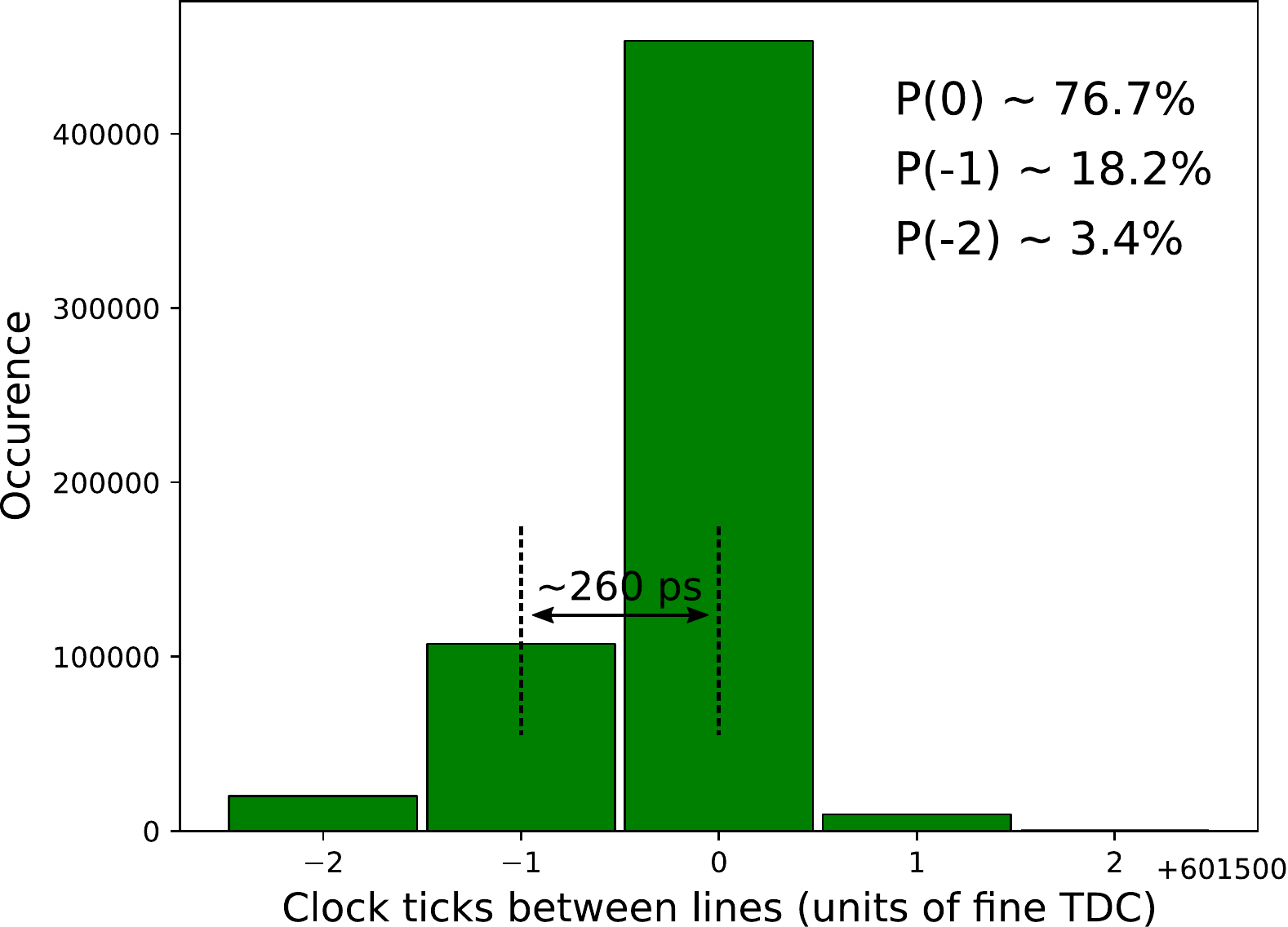}
    \caption{Histogram representing the number of clock ticks measured by the TDC in the TPX3 between successive scan rows. The average drift between our SU and the TXP3 clock is 1 cycle ($\sim$ 260 ps) every 2615217 clock ticks. The clocks can be synchronized with every scan row or every complete frame.}
    \label{linehisto}
\end{figure}

\clearpage

\section*{EELS spectra dependence on the total cluster ToT}

\begin{figure}[H]
    \centering
    \includegraphics[width=0.85\textwidth]{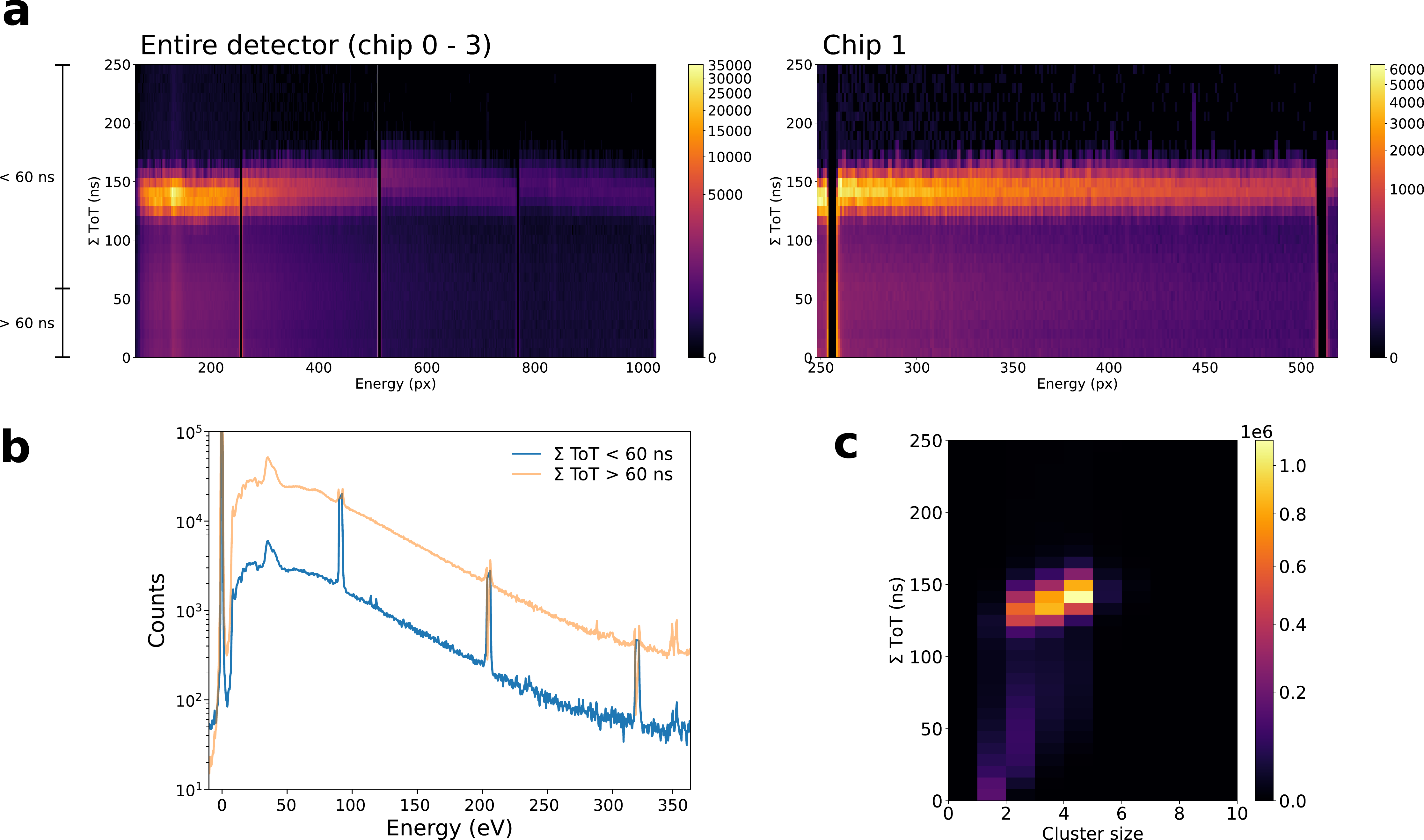}
    \caption{Relation of the summed ToT with the electron hits in the dispersive direction. (a) For the entire detector (chip 0 - 3) but excluding zero loss for a better contrast, and for the chip 1, in which the low sum ToT can be more easily identified within the chip borders. (b) The accumulated resultant spectra for the two regions identified as $\Sigma$ ToT $>$ 60 ns and $<$ 60 ns. For the low ToT, the hit count between chips is orders of magnitude greater than neighbor pixels. (c) The relation between the summed ToT and the cluster size, displaying an average of 2.92 electrons/cluster.}
    \label{supClusterSpec}
\end{figure}

For the same data studied in Figure \ref{fig2}, we have plotted the dependency of the summed ToT on the detector hit position in the dispersive direction in Figure \ref{supClusterSpec}, intending to clarify the origin of the Poisson distribution after filtering data by the summed ToT. The histogram in Figure \ref{supClusterSpec}a clearly shows a central ToT value around $\sim$ 139.12 ns (intense horizontal line), which is associated with the 100 keV electrons. A Gaussian fit provides a fwhm of $\sim$ 22.65 ns which is below the unit clock tick of 25 ns. The histogram of the chip 1 alone provides a better contrast to see the high number of clusters with $\Sigma$ ToT $<$ 60 ns around the chip junctions (close to pixels 256 and 512), Tis region, additionally, is also known to have hot pixels with high counting rates even in the absence of the electron beam. In figure \ref{supClusterSpec}b we have plotted the accumulated EELS spectrum for the lower (blue) and upper (orange) sides of the aforementioned histogram, which displays a disproportionate number of counts in the chip boundary, in the blue curve. Finally, we show in Figure \ref{supClusterSpec}c the relation between the summed ToT and the cluster size, which has its maximum of around 139.12 ns for a cluster size of 4 electrons. Note also the high number of clusters with a unit cluster size (individual electrons) for lower ToTs, which can be caused by the physical boundary condition on-chip junctions.

\clearpage

\section*{Cluster-detection algorithm results for different parameters}

\begin{figure}[H]
    \centering
    \includegraphics[width=0.85\textwidth]{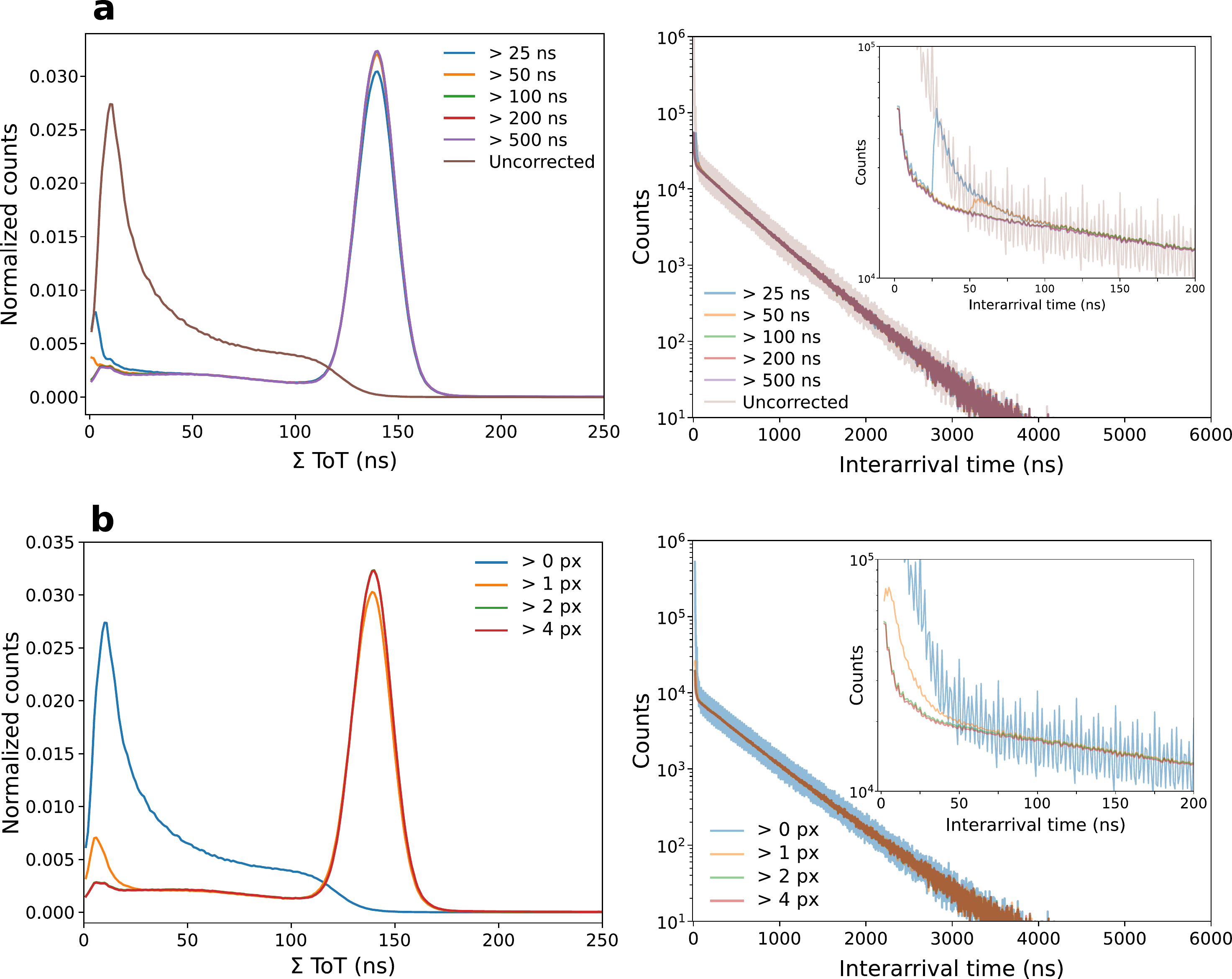}
    \caption{Results of the cluster-detection algorithm for different identification parameters. (a) Variation of $\Delta$ToA using $\Delta \alpha$ = $\Delta \beta$ = 2. Note how 25 ns and 50 ns underestimate cluster formation time. $\Delta$ToA $>=$ 200 ns is enough for the algorithm convergence. (b) Variations of $\Delta \alpha$ = $\Delta \beta$ using $\Delta$ToA = 200 ns. Convergence is achieved for values $>=$ 2, meaning that we must have at least 3 pixels of distance (in each $\alpha$ and $\beta$ independently) in order to produce a new cluster.}
    \label{supClusterPar}
\end{figure}

As briefly discussed in the main text, cluster detection is performed by sorting the pixel hits by ToA and comparing them neighbor by neighbor: if the ToA increases by more than $\Delta $ToA or if the pixel spatial position increases by more than either $\Delta \alpha$ or $\Delta \beta$, a new cluster is formed. This procedure is performed for hits that are not yet within a cluster until convergence is achieved. Figure \ref{supClusterPar}a shows the same plot as in main text (Figure \ref{fig2}) for different $\Delta$ToA values and for the uncorrected data set using $\Delta \alpha$ = $\Delta \beta$ = 2. As the pixel dead time is $\sim$ 475 ns, there is no risk of same-pixel cluster identification as long as $\Delta$ToA is smaller than this value. The typical cluster formation interval, however, is smaller than 475 ns and one must be primarily careful not to underestimate $\Delta$ToA. This is undoubtedly the case for 25 ns, 50 ns, and, to a lesser extent, 100 ns. $\Delta$ToA of 200 ns and 500 ns provide almost identical data. In Figure \ref{supClusterPar}b, we have varied $\Delta \alpha$ and $\Delta \beta$ equally within the range $[0, 4]$ for $\Delta$ToA = $200$ ns. For $\Delta \alpha$ = $\Delta \beta$ = 0, the result is identical to that for the uncorrected data, which is obvious considering any electron detected within $\Delta$ToA = 200 ns will be at a different pixel due to the pixel dead time. Data does converge for $\Delta \alpha$ = $\Delta \beta$ $>=$ 2.

Note that there is no conflict in choosing $\Delta \alpha = \Delta \beta = 2$ and the cluster size shown in Figure \ref{supClusterSpec}c. Indeed, one can have clusters of 5-10 electrons as long as the pixel separation between consecutive electrons in the ToA-sorted list is $<= 2$. For example, consider the following pixel address electron events list, in which their ToA is less than 200 ns between each other: (128, 128), (129, 128), (130, 128), (132, 130), (132, 132). There is a maximum separation of 2 pixels between consecutive electron events in each address direction and thus they belong to the same cluster. In this case, the cluster size is 5.

\clearpage




\end{document}